\documentclass[preprint, 3p, twocolumn, 12pt]{elsarticle}

\usepackage{amssymb}
\usepackage{array}
\usepackage{caption}
\usepackage{subcaption}
\usepackage{enumitem}
\usepackage{float}
\usepackage{xcolor}
\usepackage{multirow}

\usepackage[hyphens]{url}
\usepackage[hidelinks]{hyperref}
\hypersetup{breaklinks=true}

\usepackage{booktabs}% http://ctan.org/pkg/booktabs

\journal{Automation in Construction}
\begin{document}
\begin{frontmatter}

% \title{Converging Design Automation: Exploring the Overlap Between Electronics and Building Design}
\title{From Electronic Design Automation to Building Design Automation: Challenges and Opportunities}
% \title{Synthesis of Building Design: An Overview}
% \title{Toward Design Automation in the Building Design Process}
% \title{Design Automation in the building design process: An Overview}
% \title{Toward Defining Level of Abstraction for Building Energy Models}

% 5000-7000 words suggested



\author[inst1]{Yu-Wen Lin}
% \author[inst2]{Aariz Hazari}
\author[inst2]{Tsz Ling Elaine Tang}
\author[inst1]{Alberto L. Sangiovanni-Vincentelli}
\author[inst3]{Stefano Schiavon}
\author[inst1]{Costas J. Spanos}

\affiliation[inst1]{organization={Department of Electrical Engineering and Computer Sciences, University of California, Berkeley},
            % addressline={Address One}, 
            city={Berkeley},
            % postcode={00000}, 
            state={CA},
            country={USA}}
% \affiliation[inst2]{organization={Department of Mechanical Engineering, University of California, Berkeley},
%             % addressline={Address Two}, 
%             city={Berkeley},
%             % postcode={22222}, 
%             state={CA},
%             country={USA}}

\affiliation[inst2]{organization={Siemens Technology},
            addressline={755 College Rd East}, 
            city={Princeton},
            postcode={08540}, 
            state={NJ},
            country={USA}}

\affiliation[inst3]{organization={Center for Built Environment, University of California, Berkeley},
            % addressline={55 College Rd East}, 
            city={Berkeley},
            % postcode={08540}, 
            state={CA},
            country={USA}}
\begin{abstract}
Design automation, which involves the use of software tools and technologies to streamline the design process, has been widely adopted in the electronics industry, resulting in significant advancements in product development and manufacturing.
However, building design, which involves the creation of complex structures and systems, has traditionally lagged behind in leveraging design automation technologies. Despite extensive research on design automation in the building industry, its application in the current design of buildings is limited.
This paper aims to (1) compare the design processes between electronics and building design, (2) highlight similarities and differences in their approaches, and (3) examine challenges and opportunities associated with bringing the concept of design automation from electronics to building design.
\end{abstract}


\end{frontmatter}


\section{Introduction}
\label{sec:intro}
% Outline
% Design automation definition and how it is currently used in various industries
% Overview of the building design process and its potential for automation
% Importance of design automation
Design automation refers to the utilization of software tools to aid designers in creating and delivering products more efficiently by eliminating the need for manual labor that was required with traditional design methods.
It has gained prominence in several industries, including electronics \cite{macmillen2000industrial}, automotive \cite{lan2018design}, aerospace \cite{shmelova2019automated}, and manufacturing \cite{lu2020smart}, where it has enabled significant advancements in product development and manufacturing efficiency.
However, in the building industry, design automation has not been adequately utilized and has yet to become a standard part of the design process. 

The building design process involves several parties, including architects, engineers, building authorities, clients, financial institutions, and constructors. 
Traditionally, different development disciplines are siloed from each other and each section is assembled after the product is set in stone. 
However, this separate independent design results in not only personnel dependencies that increase the time of the critical path from initial design to complete development but also in objective inefficiencies that result from a lack of communication during the development process. 
Stronger communication between all the stakeholders would be essential to improve the development process for energy-efficient buildings.

Reed et al. \cite{reed2009integrative} developed an integrative building design framework to guide involved parties to communicate more efficiently. 
Leadership in Energy and Environmental Design (LEED), a green building rating system, also promotes an integrative process by providing one point in credit to encourage synergies between different disciplines during the design process. 
Though Building Information Modeling (BIM) and the integrative framework exist for more than a decade, the current building industry still has yet to adapt to the integrative design process. 
Challenges include integration between different software and tools, building complexity, communication barriers between different design teams, and limited standardization for building design practices. 

The building design process shares several similarities with the electronic design process. 
They both require an initial concept or idea to be developed and refined through multiple stages of design and prototyping.
In both cases, the design must consider various technical requirements, such as functionality, safety, and durability.
Additionally, both building and electronic design require collaboration among different professionals, including engineers and designers. 
Furthermore, both design processes often involve the use of specialized software and tools to aid in the creation and testing of the design.
However, while the electronic industry has evolved from hand-drawn design to Electronic Design Automation (EDA), the building industry has yet to fully utilize design automation technologies.

Although there has been research on design automation techniques in subsystems of building design, such as structural form \cite{yi2009optimizing}, floor plan \cite{de2016genetic, wong2009evoarch, feng2016crowd}, fa\c{c}ade \cite{wang2007facade, torres2007facade, newton2019generative}, and Heating, Ventilation, and Air Conditioning (HVAC) \cite{barnaby2001hvac, zhang2005synthesis},  these techniques are rarely applied in  the current building industry. 
One of the main reasons is the absence of standardized input-output relationships for each component.
This lack of standardization makes it challenging to reuse components across different designs. 
Consequently, it becomes difficult to remove and replace a specific component or technique in another design, as there is no assurance that the inputs and outputs will align correctly.
Additionally, each building type has its unique requirements, constraints, and regulations, resulting in a tailored design process for each specific type and location.
However, automation can provide several benefits to the building design process, such as increased efficiency, improved accuracy, cost reduction, consistency, flexibility, and innovation. 

By leveraging automation, designers can complete tasks more efficiently and accurately, reduce errors and inconsistencies, reduce or eliminate manual labor, ensure that design standards and procedures are consistently followed, enable easy modification and adjustment of designs, and experiment with new and innovative design ideas.
In this paper, we will use EDA as a reference to identify the areas where building design falls short in realizing automation and explore the opportunities that can arise if automation is implemented.
We aim to (1) compare and contrast the design process of buildings and electronics, (2) identify challenges that hinder the automation of building design, and (3) provide future research directions to realize design automation in the building industry.

\section{Building Design}
% Outline
% 1- Stages of building design
% 2- Complexity of building design (challenges?)
% 3- Negative consequences of poor building design

% Physical design is tedious and error-prone
\begin{figure*}[]
    \centering
    \includegraphics[width=\textwidth]{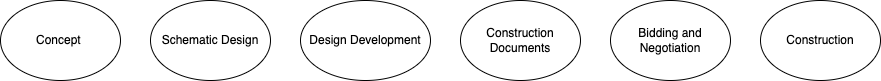}
    \caption{Building design process}
    \label{fig:iterative}
\end{figure*}

Building design has a significant impact on energy consumption and the worldwide carbon footprint.
The following sections discuss the state-of-the-art building design process, common software tools used in building design, ongoing automation efforts in research, and the negative consequences of poor building design.

\subsection{Design Work Flow}
% Building involves several stakeholders
The building work flow is a complex and iterative process that involves various stages and stakeholders. 
It is generally split into six different stages: concept, schematic design, design development, construction documents, bidding and negotiation, and construction as shown in Figure \ref{fig:iterative}.
The absence of an arrow connecting to each node suggests that the building design process is not a linear left-to-right process but rather an iterative one.
This means that the design goes through several stages and can backtrack to previous stages for refinement at any point in the process. 
For example, during the construction of a building that may take several years, the building codes or regulations may change, necessitating design changes in the early design stages. 
Therefore, the design process requires a flexible approach to accommodate changes in requirements and constraints, ensuring that the final design meets the necessary standards and regulations. 

The concept design stage involves gathering information about the project's goals, site, budget, and other constraints. 
It may also include conducting site analysis and surveys and reviewing zoning regulations.
The schematic design involves developing preliminary design concepts, sketches, and drawings that capture the client's requirements and vision.
Generally, the building's form, size, and general layout are defined at this stage.
In the design development stage, the preliminary design is further developed, refined, and detailed. 
It includes more detailed drawings, specifications, and material selections.
During the construction documents stage, all the design documents are prepared into detailed construction documents that include detailed drawings, specifications, and instructions for the contractors to follow.
After completing the construction documents, the next step entails soliciting bids from contractors and engaging in contract negotiations.
Lastly, the construction stage involves overseeing the construction process to ensure that the project is built according to its design and specifications.

Throughout the process, the building designers will work with various stakeholders, including the client, engineers, contractors, and government officials, to ensure that the project meets the client's requirements, complies with regulations and standards, and is constructed safely and efficiently.
The process is often iterative, with the designer revising the design as new information and feedback are received.

\subsection{Building Design Software Tools}
The building design process relies heavily on software tools to aid in the creation, management, and execution of design projects. 
Some of the most common software tools used in building design include Computer-Aided Design (CAD) software, Building Information Modeling (BIM) software, energy modeling software, project management software, and other analysis software.
CAD software is used to create detailed 2D and 3D designs of building components. 
BIM software is a collaborative tool that allows architects, engineers, and contractors to work together in an integrated platform. 
This software provides a shared space for design and construction data, enabling all parties to view, edit, and comment on the project in real-time.
The most common ones are Autodesk Revit \cite{revit}, ArchiCAD \cite{archicad}, and Bently OpenBuildings Designer \cite{bently}.
There are various data structures used to store information in BIM software.
Luo et al. \cite{luo2021datatools} have summarized 24 different data tools for representing and managing building data.
These data tools help to organize and structure information in the BIM, enabling more efficient and accurate design, construction, and maintenance processes.
Energy modeling software plays a crucial role in simulating building energy use and identifying opportunities for energy optimization and savings.
By using these tools, designers can make informed decisions about building materials, heating and cooling systems, and lighting to improve energy efficiency and reduce costs.
The most common energy simulation software tools for the building include EnergyPlus \cite{crawley2001energyplus}, IESVE \cite{iesve}, and eQUEST \cite{eQuest}.
Attia et al. \cite{attia2011early} also conducted a comparative analysis of ten common energy simulation tools that are typically used in the early design stages of buildings. 
Project management software is used to manage the building design process from start to finish. 
This software allows project managers to create schedules, assign tasks, and track progress, ensuring that projects are completed on time and within budget.
Other analysis software is also used to assess and optimize various aspects of building design, such as structural analysis, acoustics, and lighting. 
However, there are limitations in existing software tools.
One major issue is their complexity, which can require significant training and expertise to use effectively, making them inaccessible to small firms or individual designers. 
Integration issues can also arise when working with different software tools, leading to errors and inconsistencies when data is manually transferred between platforms.
The accuracy of simulation software can also be limited by incomplete or inaccurate data. 
Despite these challenges, software tools remain essential for creating high-quality, efficient, and cost-effective building designs.
By leveraging these tools, designers can improve collaboration and communication among team members, streamline workflows, and ultimately create better building designs.

\subsection{Ongoing Design Automation Efforts}
Traditional design approaches are often limited by the experiences and knowledge of human designers.
Key design decisions are often already made before assessment due to budget, time constraints, and project requirements.
In addition, design evaluations primarily serve to meet building codes such as energy standards, ASHRAE 90.1 \cite{halverson2014ansi}, and thermal environmental conditions, ASHRAE 55 \cite{ashrae55}.  
The building design has the potential to be optimized further to create a more energy-efficient and environmentally-friendly solution.
Advances in computing technologies and machine learning algorithms have enabled the possibility of exploring a larger design space that may be overlooked by designers.
For example, Generative Adversarial Networks (GANs) can be used to generate new building designs \cite{sun2022automatic, wu2022generative}, Genetic algorithms (GA) can be used to optimize building parameters and configurations with an objective of energy efficiency and thermal comfort \cite{tuhus2010genetic, li2017genetic}, and Natural Language Processing (NLP) algorithms can analyze building codes and regulations \cite{zhang2021deep, ding2022applications}. 
% https://www.autodesk.com/solutions/generative-design/architecture-engineering-construction#:~:text=What%20is%20Generative%20Design%20in,for%20data%2Ddriven%20decision%20making
Existing BIM software, Autodesk Revit 2021 \cite{autodesk2021}, has the capacity to implement Generative Design (GD) for building geometry. 
The feature allows users to input design goals and constraints and then uses GD to generate multiple design options to meet those goals and constraints.  
It can be used to optimize building designs for structural performance, energy efficiency, and sustainability. 
However, limitations exist in the user interface, data availability, designer bias, complex structures, and processing speed. 

Design automation is primarily present in research and has yet to be applied in the building design process in the industry. 
The main challenge is that most studies are specific and cannot be applied to all building types. 
Although complete automation of the entire design process may not be feasible due to the complexity of buildings, incorporating automation tools can streamline the design process, reduce errors, and improve efficiency and reliability. 
Utilizing advanced computational tools, designers can generate a set of high-quality solutions that meet design objectives and constraints, such as energy efficiency, functional performance, and cost-effectiveness.

% Negative consequences for poor design 
\subsection{Negative Impacts of Poor Building Design}
Poor building design can have significant negative impacts on the building occupants, the environment, and the economy.
One of the most immediate and visible consequences of poor building design is the impact on occupants' health and comfort \cite{singh1996impact, evans1998buildings, boyce2010impact, boubekri2008daylighting}. 
Poor ventilation, lighting, acoustics, and thermal comfort can all contribute to health problems such as allergies, asthma, headaches, and stress. 
These problems can lead to decreased productivity and reduced quality of life for building occupants.
Another negative impact of poor building design is its effect on the environment. 
Buildings are responsible for more than 40\% of energy consumption worldwide \cite{e_comp} and 17\% of total global greenhouse gas (GHG) emissions from operations and the manufacture of building materials \cite{lamb2021review}, and poor design can exacerbate these problems \cite{clarke2007energy}. 
For example, buildings that are poorly insulated, have inefficient heating and cooling systems, or use energy-intensive materials can contribute to higher energy consumption and carbon emissions. 
Another example is that the HVAC system in a building is often oversized due to design safety factors and a lack of accurate building load calculations \cite{djunaedy2011oversizing}.
This not only contributes to global climate change but can also lead to higher energy costs for building owners and occupants.
Lastly, poor building design can have negative economic impacts \cite{wang2014sustainable}. 
Inefficient buildings can lead to higher energy bills and maintenance costs, reducing the economic value of the building.
Poor building design can have far-reaching negative impacts on human health and comfort, the environment, and the economy.
It is essential to prioritize good design practices, including the incorporation of automation tools, to create buildings that are efficient, sustainable, and safe for their occupants.
\section{Electronic Design Automation}
% definition of EDA

While looking at the present status of the design work flow for buildings, it is instructive to compare it to that of the Electronic Design Automation (EDA) referring  to the use of software tools and algorithms to automate various aspects of designing and manufacturing complex electronic components, including today's state of the art Integrated Circuits (ICs).
EDA involves the use of computer programs that assist designers in creating, verifying, and testing electronic systems, ranging from individual circuits to complex integrated circuits and systems.
Design automation aims to improve the efficiency and productivity of the design process by reducing the time and effort required for manual tasks and by enabling designers to explore more design options and trade-offs in a shorter amount of time.
The following section describes the design process of electronics, the evolution of the EDA, and the current state of EDA and its impact.
% Outline
% a) ED process
% b) Evolution of EDA tools and their capabilities
% c) Current state of EDA and its impact on the electronics industry

% Design has moved from transitor level -> gate level -> register-transfer level -> system level 

\subsection{Electronic Design Work Flow}
\begin{figure*}[]
    \centering
    \includegraphics[width=\linewidth]{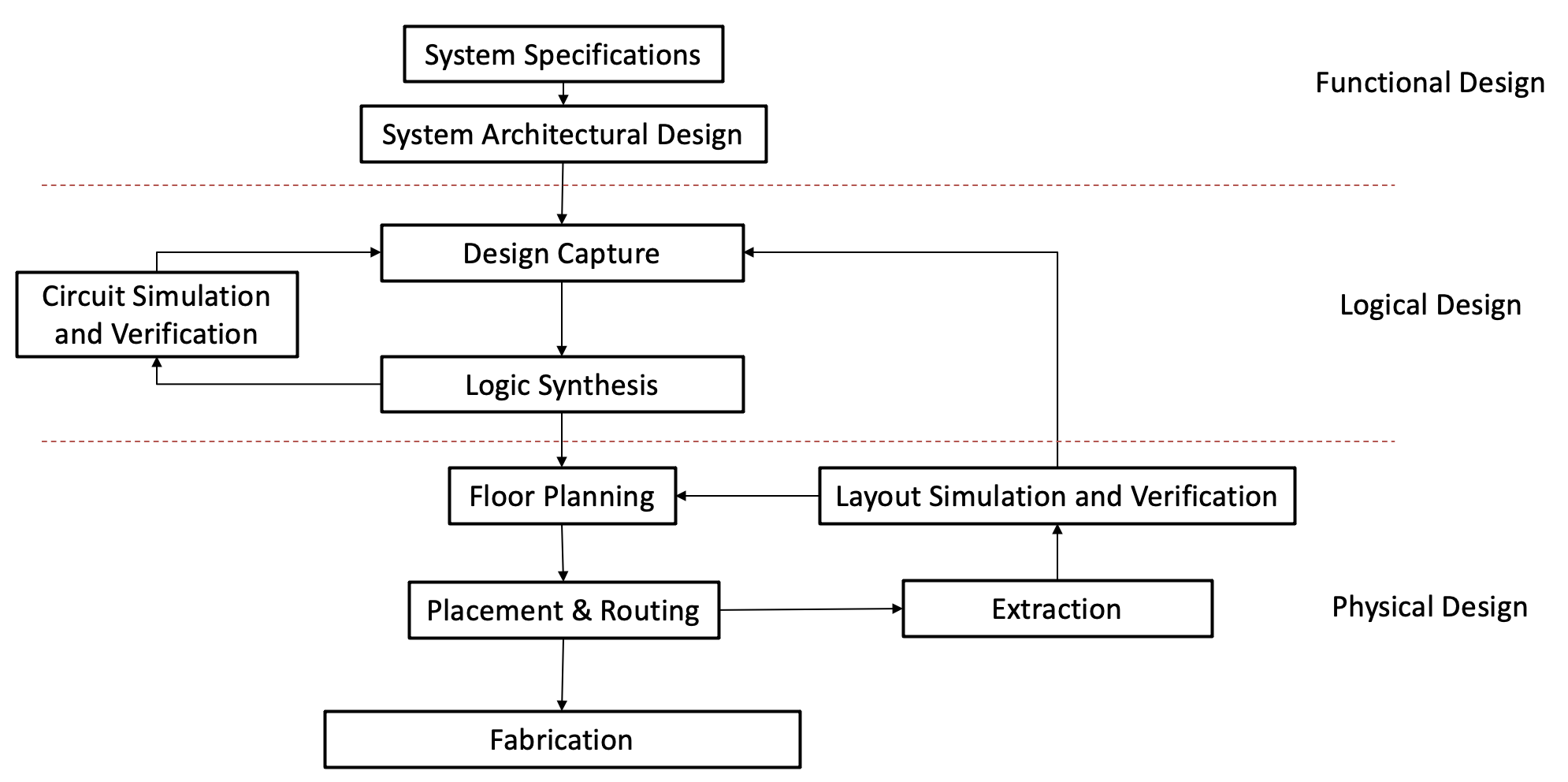}
    \caption{Electronic Design Process}
    \label{fig:edp}
\end{figure*}

The process of electronic design can vary depending on the project size and specifications, but generally, for an IC, it follows a top-down design process \cite{Browy2014} as shown in figure \ref{fig:edp}.
The first step is to define the system specifications, which include the desired performance, functionality, and power consumption of the system.
Based on these specifications, a high-level block diagram is created to define the system-level architecture and interconnections between the functional components. 
Once the architecture is defined, the next step is design capture, which involves creating a high-level design description of the system using a Hardware Description Language (HDL). 
This is followed by logic synthesis, where the captured design is transformed into a gate-level netlist that describes the circuit's logic in terms of logical gates, flip-flops, and other standard cells.
Next, circuit simulation and verification are performed to verify the circuit's functionality.
This can involve some iteration to make sure that the simulated results meet the desired specifications.
The next level of design is the physical design. 
Floor planning involves determining the physical location of the various functional blocks in the design.
The placement and routing step follows where the physical locations of individual gates and components are determined and the connections between them are established by routing wires.
After placement and routing, the circuit layout is extracted and used for simulation and verification. 
This involves ensuring that the physical layout matches the intended design and verifying the functionality and performance of the circuit. 
Any issues found during this step may require adjustments to the design, placement, or routing before final verification and fabrication.
Lastly, the design is ready for fabrication, where it is physically manufactured using the digital description of the final design.
The above process emphasizes that electronic design is an iterative, yet highly streamlined process aimed at ensuring that the functionality meets the desired specifications. 
Moreover, each design phase represents a distinct and well-defined level of abstraction, enabling effective analysis.

As technology continues to advance, electronic design has become more complex, resulting in the emergence of new design protocols. One such process is the System-on-a-Chip (SoC) design \cite{martin2001system}, which involves integrating multiple components onto a single chip.
This process requires specialized design tools and methodologies, including co-design, Intellectual Property (IP) reuse, and system-level verification, to ensure that the individual components work together effectively.
Another emerging design process is 3D IC design \cite{kondo2015three}, which involves stacking multiple layers of chips vertically to increase performance and reduce size. 
This process involves several new challenges, including thermal management, power distribution, and signal integrity.
However, for the purpose of this paper, we will mainly focus on the top-down approach in electronic design, which is more similar to the current building design work flow.

\subsection{Evolution of EDA}
\begin{figure*}[]
    \centering
    \includegraphics[width=\linewidth]{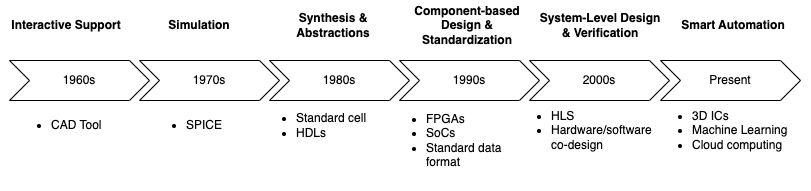}
    \caption{Timeline of Key Developments of EDA}
    \label{fig:timeline}
\end{figure*}
EDA has a significant impact on the semiconductor industry by enabling the faster, more efficient, and more reliable design of electronic systems.
EDA tools provide designers with a range of features and capabilities that simplify and automate the design process, enabling them to create complex designs more quickly and accurately.
Fig \ref{fig:timeline} shows the timeline of key developments of EDA from interactive design support to design automation.
The development of EDA tools began in the 1960s when researchers developed the first computer-aided design (CAD) tools \cite{sutherland1964sketch} designed to assist engineers in analyzing and designing circuits and boards.
It transits from a traditional hand-drawn design to a digital design.
Later, SPICE (Simulation Program with Integrated Circuit Emphasis) emerged in the early 1970s \cite{vladimirescu1980spice} as a student project, but became an industry standard for IC Design.

In the late 1970s and early 1980s, two major technologies emerged: standard cell design and hardware description language (HDLs) \cite{shiva1979computer}.
The concept of standard cells involves pre-designed circuit blocks that contain a specific logic function, such as an AND gate or a flip-flop. 
These standard cells can be combined to create more complex circuits. 
The introduction of standard cells separates the transistor-level design from the logic design. 
On the other hand, HDLs, such as VHDL and Verilog, provide a high-level language for describing digital circuits, enabling designers to specify the behavior of a circuit without having to worry about the details of the underlying hardware. 

In the 1990s, EDA tools continued to evolve, with the introduction of new tools and methodologies such as field-programmable gate arrays (FPGAs) \cite{kuon2008fpga} and system-on-chip (SoC) design \cite{martin2001system}. 
FPGAs were programmable devices that could be used to implement custom logic functions, while SoC design involved the integration of multiple functions onto a single chip.
Furthermore, the industry developed standard ASCII formats, Library Exchange Format (LEF) and Design Exchange Format (DEF) \cite{LEFDEF}, for representing library and design data in order to share and reuse design data across different tools and platforms. 

In the 2000s, High-Level Synthesis (HLS) and hardware/software co-design are introduced. 
HLS allowed designers to describe hardware using a high-level programming language, which could then be synthesized into hardware.
Hardware/software co-design involved the simultaneous design of hardware and software components, enabling the design of more complex systems that integrated both hardware and software. 

\subsection{Current State of EDA and its Impact}
The current state-of-the-art in EDA includes advanced technologies that enable designers to create complex semiconductor chips with high performance, high reliability, and low power consumption.
High-level synthesis technology \cite{gajski2012high} allows designers to specify chip functionality at a higher level of abstraction, which reduces the time and effort required for manual coding.
Machine Learning methods \cite{huang2021machine} are also applied in EDA industry to optimize design parameters to further improve design quality.
Furthermore, cloud computing \cite{sivacadence} allows designers to access on-demand computing resources to perform tasks that required high computational power without having to invest in expensive hardware and software infrastructure.
Advanced verification and validation techniques \cite{drechsler2004advanced}, such as formal verification, simulation, and emulation ensured the reliability and correctness of complex designs. 
Lastly, as chips become smaller and more complex, the placement and routing of components on the chip become more challenging for EDA tools. 
As a result, placing and routing algorithms \cite{sherwani2012algorithms} are constantly evolving to incorporate new design methodologies that can help designers to optimize chip layouts for power, performance, and manufacturability. 

Automation is necessary to handle the immense complexity of semiconductor chips, which can have over a billion circuit elements that can interact in subtle ways.
Manufacturing variations can introduce further complexities and changes in circuit behavior.
Errors in manufactured chips can be disastrous and often cannot be fixed. 
Re-designing and Re-manufacturing the entire chip is time-consuming and expensive, which can lead to project failure.
EDA technology provides the critical automation needed to manage the complexity, making it possible to design and manufacture modern semiconductor devices with a high level of accuracy and efficiency.

\section{Discussion}
\begin{figure}[H]
    \centering
    \includegraphics[width=\linewidth]{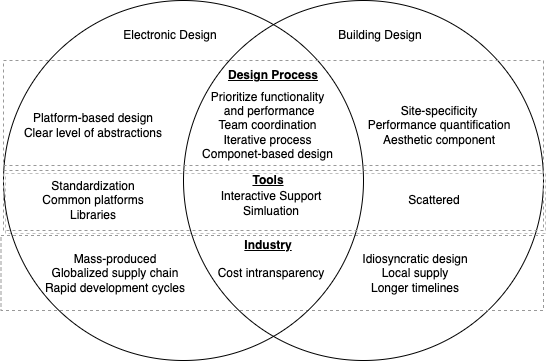}
    \caption{Comparison between electronic design and building design.}
    \label{fig:comparison}
\end{figure}

While the current building industry has yet to reach a high threshold for design automation, we can learn from the development of electronic design automation to identify the key aspects needed to make it a reality.
Figure \ref{fig:comparison} outlines the comparison between electronic design and building design.
We will examine three key aspects: design process, design tools, and industry, and compare the similarities and differences between them in greater detail in the following sections.

\subsection{Design process}
Both electronic design and building design begin with specifying the project requirements and constraints, which usually include functionality and performance. 
Additionally, both processes involve the contribution of multiple team players to complete the design. 
In electronic design, this may include designers, engineers, and other specialists, while building design usually involves architects, engineers, consultants, contractors, and constructors.
Effective collaboration between stakeholders is essential to achieving an optimal design. 
Both processes are iterative, where certain stages are repeated until the optimal design is achieved. 
The iterative process allows designers to refine and optimize the design over time, making adjustments to accommodate changing requirements or constraints.
This also helps identify potential issues early on, reducing the risk of errors in later stages of the design.
Furthermore, both design processes involve component-based design, where components are selected and integrated into the design as a whole.
In building design, this includes various building components, such as walls, windows, doors, and HVAC components, while in electronic design, it involves components such as sensors and devices depending on the level of abstraction. 

However, building design often employs a component-based design approach at the subsystem level, such as for HVAC or structural component design.
This approach is rarely available at the system level, primarily due to the lack of a common platform in the building design domain. 
Design and simulation tools used in building design are often incompatible with each other, making it difficult to integrate designs into a single platform. 
This lack of integration can lead to errors, duplication of effort, and increased design time.
The current efforts in BIM have helped to address some of these challenges by providing a platform for integrating various building design tools and facilitating the exchange of information between stakeholders. 
However, there are still limitations in software compatibility and data management.
Furthermore, the lack of clear definitions for different development objectives and levels of abstractions during different design stages can further complicate the design process. 
In contrast, the electronic industry benefits from clear and standardized input-output requirements.

\begin{table}[htbp]
\caption{The LoD for BIM \cite{lod_AIA}.}
\centering
\begin{tabular}{|m{0.2\linewidth}|m{0.7\linewidth}|}
\hline
 \textbf{LoD}	& \textbf{Model Descriptions} \\
\hline
LoD 100  & Approximate graphical representation of the entire building based on spatial requirements \\
\hline
LoD 200 &  Determine approximate floor plans and space boundaries (walls, elevation, and columns)  \\
\hline
LoD 300 & Detailed design for actual construction\\
\hline
LoD 400 & Precise model for fabrication, structure, electrical, mechanical, and plumbing systems\\
\hline
LoD 500 & Complete design, which represents the actual building\\
\hline
\end{tabular}
\label{table:loa}
\end{table}

The existing levels of detail (LoD) framework of BIM developed by the American Institute of Architects (AIA) \cite{lod_AIA} is defined in Table \ref{table:loa}.
LoD defines the amount and degree of building information that needs to be in the BIM at different stages.
It is developed for construction purposes and is not directly applicable to all disciplines during the design stages. 
The broad definition requires elaboration to fit into design purposes to allow the development of design automation in the building industry.
The building design is often limited by site-specificity, which can make it difficult to develop a clear design process.
Each building project has unique site requirements, such as topography, climate, and local regulations.
These factors can have a significant impact on the design decisions and may require adjustments to the design throughout the process.
As a result, the design process in building design can be more complex and time-consuming than in electronic design, where products can be developed independently of a specific site
% Without a clear definition of the building design process and connection between different disciplines, researchers develop building design methodologies with various inputs and outputs based on the convenience of data availability and program capabilities.
% This not only makes it challenging to compare and contrast different design methods but also impedes the standard data flow process from one software to the next.

Another difference in the design process between electronics and buildings is the performance quantification aspect.
In electronic design, the performance metrics can be easily obtained through calculations in timing and energy performance.
However, in the building industry, several aspects need to be considered, such as energy efficiency, indoor air quality, visual comfort, thermal comfort, acoustic lighting, and water efficiency.
These aspects are not always straightforward to quantify, resulting in discrepancies among designers.
Additionally, due to various constraints, design tradeoffs become more complex and harder to make in building design.

Lastly, the level of attention to the occupant experience and the importance of aesthetics of building design are vastly different from electronic design.
Unlike electronic design, building design has to take into account the physical environment and how the building will fit into the surrounding landscape.
Designers have to consider the impact of the building on the environment, the local culture, and the community it will serve.
In contrast, electronic design focuses almost entirely on functionality and efficiency.
While the aesthetic portion of building design is challenging to fit into design automation applications, other areas such as structural design and construction planning can benefit greatly from automation.

\subsection{Design Tools}
Simulation tools are commonly used in both electronic design and building design.
These tools can provide interactive support and simulation capabilities to help designers explore design options, identify potential problems, and optimize design solutions.

In electronic design, simulation tools are highly standardized, with well-established input-output relationships that enable designers to quickly and easily model electronic circuits, test different design scenarios, and verify design performance.
There is also a common platform for simulation tools, which allows designers to use different simulation tools with similar syntax, enabling the transfer of simulation models between different software programs.

In contrast, building design lacks a universal standard for simulation tools, which can make it challenging for designers to choose the right tool for their specific design problem.
Energy simulation software, for example, may require different inputs for different programs, making it difficult to have a common platform for simulation tools.
Furthermore, there are limited data standards that cover all simulation tools, and data conversion between tools often requires manual adjustment.
Building designers or engineers often need to adapt to different software programs, such as structure, energy, and lighting simulations, for different design purposes, which results in a fragmented design process.
Interoperability is a widely known issue in the building industry \cite{Hong_buildsim_challenges}.
% IFC, MVD, Brick Schema
Data standards, such as Industry Foundation Class (IFC) \cite{IFC_iso201816739} and Green Building XML (gbXML) \cite{gbxml}, exist in the building industry.
IFC is used to exchange building data, including geometry, spatial elements, and sensor information.
A subset of IFC, Model View Definition (MVD), has been developed to apply to specific applications in building \cite{hietanen2006ifc}. 
It defines which IFC entities, attributes, and relationships are required to support a specific use case or task. 
The gbXML format is more specifically used in the energy simulation domain. 
It facilitates the exchange of data between CAD and energy analysis tools. 
Other data tools in the building industry are summarized in \cite{luo2021datatools}.
However, the above-mentioned data standards have yet to be applied to all building applications.
Extensions or adaptations are still required to apply to specific building applications or regions \cite{froese2003future}.

On the other hand, electronic data standards, LEF and DEF \cite{LEFDEF}, are well-established in exchanging data between different software programs.
LEF is used to exchange library data, while DEF is used to exchange design data. 
Building data standards need to have broad adoption across the industry and cover all relevant disciplines.
Different from LEF and DEF, it needs to be designed to be flexible and adaptable to different use cases and applications in building.
Moreover, a library data standard needs to exist to store different application modules within the building industry.
Its primary purpose is to enable seamless data exchange between different software applications used in building design, construction, and operation.
By doing so, it enables the industry to create a shared library that allows for efficient sharing and reusing of data for future design processes.

In addition to the challenges of standardization and interoperability, another pivotal issue in building design is maintaining up-to-date, high-quality data throughout the design process and among stakeholders.
Failure to keep data updated can result in miscommunication, additional time required, and increased costs.
Having the data updated is especially important during the construction process, where delays and errors can have significant impacts on project time and costs \cite{zhao2019real}. 
Therefore, having an efficient and effective data management system is also vital for building design.

\subsection{Industry}
In building design, it can be challenging to estimate project costs upfront, which can lead to difficulties in accurately estimating the overall project cost.
The cost is determined through a construction bidding process. 
As a result, the cost of sub-systems is usually not publicly available.
Several researchers have explored various methods to improve cost estimation in the building industry \cite{holm2021construction}, including the use of historical cost data, and cost modeling techniques.
Similarly, in the electronic industry, cost estimation is also challenging due to the competitive nature of the industry. 
For in-house manufacturing products, the cost is generally easier to estimate because the manufacturing process is more standardized and predictable.
On the other hand, the cost of custom chip designs that are outsourced for manufacturing can be more challenging to estimate, as it may depend on factors such as the complexity of the design, the choice of manufacturing partners, and the timing of the project.

One major difference between the two industries is the nature of the products produced.
Electronic design often involves mass-produced products, while building design usually involves idiosyncratic designs. 
In electronic design, products are typically manufactured in large quantities, with standardized components and design processes, which can lead to more streamlined production and lower costs. 
Building design, on the other hand, is often unique to each project and requires more customized solutions.
As a result, the design  process needs to be flexible to accommodate a wide range of project types and requirements.

There is also a notable difference in the scale and timeline of outcomes between electronic design and building design.
In electronic design, products are often smaller in scale and require shorter development cycles due to the nature of the industry.
This allows for rapid prototyping, testing, and production, which can result in more agile and iterative design processes.
On the other hand, building design projects involve larger-scale outcomes that often require longer timelines due to the complex nature of the construction process.
This can result in a less agile design process, with greater risks and costs associated with design changes over time.

Another significant difference is the supply chain.
The electronic design industry often relies on a global supply chain, with components sourced from all over the world.
In contrast, the building design industry often relies on local suppliers and materials, which can vary depending on the region or location of the project.
This can affect the availability, cost, and quality of materials, as well as the overall design and construction process.

\section{Conclusion and Future Work}

Buildings are complicated systems that consist of different layers of sub-systems.
In addition, decisions made in the early stages of building projects have an important impact on material and energy efficiency.
It is a challenging task to frame and solve one multi-objective optimization problem for finding an optimal design solution for a building.
There is a need to develop a standardized information flow specifically for different disciplines. 
This paper provides a comparison between the electronic design and building design industries, with a focus on exploring the potential for design automation applications in building design.

Despite ongoing research efforts to streamline the building design process, the industry has been slow to adopt new technologies and approaches.
One of the key obstacles to widespread adoption is the lack of a universally defined design workflow in the industry.
Building design practices can vary significantly based on factors such as location, building type, and project scope, making it challenging for companies to invest in optimization and co-optimization techniques during the design phase.
Many optimization techniques may not be applicable to all projects, leading to concerns about cost-effectiveness and return on investment.

% 1. Standardized design process
To overcome this challenge, the industry may need to focus on developing more standardized approaches to building design, with greater emphasis on collaboration and knowledge-sharing among stakeholders.
This could include the development of a platform-based or modular-based approach that is flexible and adaptable to different project types and scopes.
Modularization involves dividing buildings into assemblies of components that have standardized interfaces for communication, making it possible to combine and reconfigure components to meet different design requirements.
This allows for a more flexible design to meet users' preferences and performance requirements. 
This approach offers several benefits, including greater flexibility in design, more efficient use of resources, and improved cost-effectiveness.

% Integration Platform/ Backbone of the design
Additionally, to facilitate more efficient and collaborative building design, it is necessary to create a design backbone that enables single-platform integration.
First, data exchange standards should be revised to ensure they are applicable to all disciplines involved in building design.
This will help ensure that information can flow smoothly between different software applications and enable effective collaboration.
Next, a Building Information Modeling (BIM) system, such as Revit, can be used as a common platform for communication between different software tools.
By integrating all existing software tools into the BIM system through support plug-in application modules, designers can access all the necessary information in one place, improving efficiency and reducing errors.

% Performance quantification
Building design requires a more comprehensive approach that considers several aspects of performance, including energy efficiency, indoor air quality, visual comfort, thermal comfort, acoustic lighting, and water efficiency.
There is a need to define and standardize a performance quantification framework for each aspect of concern in the building industry. 
While building codes and standards exist to facilitate better building designs, there is still a lack of a comprehensive framework in certain areas, such as visual comfort.
Ongoing efforts are made in the research domain.
For example, Cavieres \cite{cavieres2018functional} established a framework that allows the functional and behavioral systems of the building to interact in an interface.
This framework involves mapping the specific systems needed to calculate the relevant performance measures, which in turn helps designers make more informed design choices.
By quantifying and standardizing various aspects of building performance, designers can better understand the tradeoffs between different design options.

% Risk assessment
Lastly, risk assessment is another important aspect to address in building design.
Although innovative designs can lead to reduced energy consumption and GHG emissions, there is no guarantee that the desired outcomes will be achieved during construction.
If the design conditions cannot be met, construction costs can increase significantly, and contractors may be unwilling to take on the project.
A risk assessment module can help mitigate these risks by simulating various design scenarios and estimating the uncertainty associated with each design.
This will allow stakeholders to make informed decisions and select designs that are more likely to succeed in the construction phase.

% Summary
The implementation of standardization, performance quantification, and risk assessment modules is necessary to move building design toward automation.
Standardization promotes a more structured approach to building design, allowing for better data exchange and knowledge-sharing between stakeholders. 
By implementing performance quantification frameworks, designers can make more informed decisions regarding various aspects of building design.
Risk assessment can help mitigate potential issues during the design process, preventing costly and time-consuming delays during construction.
The use of these tools not only streamlines the design process but also leads to more sustainable building designs that meet performance and safety standards.
The ultimate goal is partial design automation, where designers and engineers can leverage technology to optimize building performance and minimize the environmental impact.
% \input{discarded/2-synthesis}
% \input{discarded/3-modularization}

% \appendix

 \bibliographystyle{elsarticle-num} 
 \bibliography{refs}

\begin{thebibliography}{10}
\expandafter\ifx\csname url\endcsname\relax
  \def\url#1{\texttt{#1}}\fi
\expandafter\ifx\csname urlprefix\endcsname\relax\def\urlprefix{URL }\fi
\expandafter\ifx\csname href\endcsname\relax
  \def\href#1#2{#2} \def\path#1{#1}\fi

\bibitem{macmillen2000industrial}
D.~MacMillen, R.~Camposano, D.~Hill, T.~W. Williams, An industrial view of
  electronic design automation, IEEE transactions on computer-aided design of
  integrated circuits and systems 19~(12) (2000) 1428--1448.

\bibitem{lan2018design}
S.~Lan, C.~Huang, Z.~Wang, H.~Liang, W.~Su, Q.~Zhu, Design automation for
  intelligent automotive systems, in: 2018 IEEE International Test Conference
  (ITC), IEEE, 2018, pp. 1--10.

\bibitem{shmelova2019automated}
T.~Shmelova, Y.~Sikirda, N.~Rizun, D.~Kucherov, K.~Dergachov, Automated Systems
  in the Aviation and Aerospace Industries, IGI Global, 2019.

\bibitem{lu2020smart}
Y.~Lu, X.~Xu, L.~Wang, Smart manufacturing process and system automation--a
  critical review of the standards and envisioned scenarios, Journal of
  Manufacturing Systems 56 (2020) 312--325.

\bibitem{reed2009integrative}
B.~Reed, et~al., The integrative design guide to green building: Redefining the
  practice of sustainability, Vol.~43, John Wiley \& Sons, 2009.

\bibitem{yi2009optimizing}
Y.~K. Yi, A.~M. Malkawi, Optimizing building form for energy performance based
  on hierarchical geometry relation, Automation in Construction 18~(6) (2009)
  825--833.

\bibitem{de2016genetic}
A.~de~Almeida, B.~Taborda, F.~Santos, K.~Kwiecinski, S.~Eloy, A genetic
  algorithm application for automatic layout design of modular residential
  homes, in: 2016 IEEE International Conference on Systems, Man, and
  Cybernetics (SMC), IEEE, 2016, pp. 002774--002778.

\bibitem{wong2009evoarch}
S.~S. Wong, K.~C. Chan, Evoarch: An evolutionary algorithm for architectural
  layout design, Computer-Aided Design 41~(9) (2009) 649--667.

\bibitem{feng2016crowd}
T.~Feng, L.-F. Yu, S.-K. Yeung, K.~Yin, K.~Zhou, Crowd-driven mid-scale layout
  design., ACM Trans. Graph. 35~(4) (2016) 132--1.

\bibitem{wang2007facade}
L.~Wang, H.~W. Nyuk, S.~Li, Facade design optimization for naturally ventilated
  residential buildings in singapore, Energy and Buildings 39~(8) (2007)
  954--961.

\bibitem{torres2007facade}
S.~L. Torres, Y.~Sakamoto, Facade design optimization for daylight with a
  simple genetic algorithm, in: Proceedings of Building Simulation, Vol. 2007,
  2007, pp. 1162--1167.

\bibitem{newton2019generative}
D.~Newton, Generative deep learning in architectural design, Technology|
  Architecture+ Design 3~(2) (2019) 176--189.

\bibitem{barnaby2001hvac}
C.~S. Barnaby, M.~Shnitman, W.~A. Wright, Hvac system design automation:
  issues, methods, and ultimate limits, in: Proc. of Building Simulation ‘01,
  7th International IBPSA Conference, 2001, pp. 1151--1157.

\bibitem{zhang2005synthesis}
Y.~Zhang, Synthesis of optimum hvac system configurations by evolutionary
  algorithm, Ph.D. thesis, Loughborough University (2005).

\bibitem{revit}
Autodesk, Revit: Bim software for designers, builders, and doers,
  \url{https://www.autodesk.com/products/revit} (2000).

\bibitem{archicad}
GraphiSoft, Graphisoft archicad, \url{https://graphisoft.com/us} (2021).

\bibitem{bently}
Bently, Openbuildings designer,
  \url{https://www.bentley.com/en/products/product-line/building-design-software/openbuildings-designer}
  (2022).

\bibitem{luo2021datatools}
N.~Luo, M.~Pritoni, T.~Hong, An overview of data tools for representing and
  managing building information and performance data, Renewable and Sustainable
  Energy Reviews 147 (2021) 111224.

\bibitem{crawley2001energyplus}
D.~B. Crawley, L.~K. Lawrie, F.~C. Winkelmann, W.~F. Buhl, Y.~J. Huang, C.~O.
  Pedersen, R.~K. Strand, R.~J. Liesen, D.~E. Fisher, M.~J. Witte, et~al.,
  Energyplus: creating a new-generation building energy simulation program,
  Energy and buildings 33~(4) (2001) 319--331.

\bibitem{iesve}
I.~E.~S. Limited, Ies virtual environment 2022,
  \url{https://www.iesve.com/software/virtual-environment} (2022).

\bibitem{eQuest}
D.~of~Energy~(DOE), equest version 3.65, \url{https://www.doe2.com/equest/}
  (2018).

\bibitem{attia2011early}
S.~Attia, A.~De~Herde, Early design simulation tools for net zero energy
  buildings: a comparison of ten tools, in: International Building Performance
  Simulation Association 2011, 2011.

\bibitem{halverson2014ansi}
M.~A. Halverson, M.~I. Rosenberg, P.~R. Hart, E.~E. Richman, R.~A. Athalye,
  D.~W. Winiarski, Ansi/ashrae/ies standard 90.1-2013 determination of energy
  savings: Qualitative analysis, Tech. rep., Pacific Northwest National
  Lab.(PNNL), Richland, WA (United States) (2014).

\bibitem{ashrae55}
Ashrae standard 55-thermal environmental conditions for human occupancy, Tech.
  rep., ASHRAE Inc. (1992).

\bibitem{sun2022automatic}
C.~Sun, Y.~Zhou, Y.~Han, Automatic generation of architecture facade for
  historical urban renovation using generative adversarial network, Building
  and Environment 212 (2022) 108781.

\bibitem{wu2022generative}
A.~N. Wu, R.~Stouffs, F.~Biljecki, Generative adversarial networks in the built
  environment: A comprehensive review of the application of gans across data
  types and scales, Building and Environment (2022) 109477.

\bibitem{tuhus2010genetic}
D.~Tuhus-Dubrow, M.~Krarti, Genetic-algorithm based approach to optimize
  building envelope design for residential buildings, Building and environment
  45~(7) (2010) 1574--1581.

\bibitem{li2017genetic}
T.~Li, G.~Shao, W.~Zuo, S.~Huang, Genetic algorithm for building optimization:
  State-of-the-art survey, in: Proceedings of the 9th international conference
  on machine learning and computing, 2017, pp. 205--210.

\bibitem{zhang2021deep}
R.~Zhang, N.~El-Gohary, A deep neural network-based method for deep information
  extraction using transfer learning strategies to support automated compliance
  checking, Automation in Construction 132 (2021) 103834.

\bibitem{ding2022applications}
Y.~Ding, J.~Ma, X.~Luo, Applications of natural language processing in
  construction, Automation in Construction 136 (2022) 104169.

\bibitem{autodesk2021}
Generative design for architecture, engineering \& construction, Autodesk
  (2021).

\bibitem{singh1996impact}
J.~Singh, Impact of indoor air pollution on health, comfort and productivity of
  the occupants, Aerobiologia 12~(1) (1996) 121--127.

\bibitem{evans1998buildings}
G.~W. Evans, J.~M. McCoy, When buildings don’t work: The role of architecture
  in human health, Journal of Environmental psychology 18~(1) (1998) 85--94.

\bibitem{boyce2010impact}
P.~R. Boyce, The impact of light in buildings on human health, Indoor and Built
  environment 19~(1) (2010) 8--20.

\bibitem{boubekri2008daylighting}
M.~Boubekri, Daylighting, architecture and health: building design strategies,
  Routledge, 2008.

\bibitem{e_comp}
A.~Allouhi, Y.~E. Fouih, T.~Kousksou, A.~Jamil, Y.~Zeraouli, Y.~Mourad,
  \href{http://www.sciencedirect.com/science/article/pii/S0959652615007581}{Energy
  consumption and efficiency in buildings: current status and future trends},
  Journal of Cleaner Production 109 (2015) 118 -- 130, special Issue: Toward a
  Regenerative Sustainability Paradigm for the Built Environment: from vision
  to reality.
\newblock \href
  {http://dx.doi.org/https://doi.org/10.1016/j.jclepro.2015.05.139}
  {\path{doi:https://doi.org/10.1016/j.jclepro.2015.05.139}}.
\newline\urlprefix\url{http://www.sciencedirect.com/science/article/pii/S0959652615007581}

\bibitem{lamb2021review}
W.~F. Lamb, T.~Wiedmann, J.~Pongratz, R.~Andrew, M.~Crippa, J.~G. Olivier,
  D.~Wiedenhofer, G.~Mattioli, A.~Al~Khourdajie, J.~House, et~al., A review of
  trends and drivers of greenhouse gas emissions by sector from 1990 to 2018,
  Environmental research letters 16~(7) (2021) 073005.

\bibitem{clarke2007energy}
J.~Clarke, Energy simulation in building design, Routledge, 2007.

\bibitem{djunaedy2011oversizing}
E.~Djunaedy, K.~Van~den Wymelenberg, B.~Acker, H.~Thimmana, Oversizing of hvac
  system: Signatures and penalties, Energy and Buildings 43~(2-3) (2011)
  468--475.

\bibitem{wang2014sustainable}
N.~Wang, H.~Adeli, Sustainable building design, Journal of civil engineering
  and management 20~(1) (2014) 1--10.

\bibitem{Browy2014}
C.~Browy, G.~Gullikson, M.~Indovina, \href{http://www.gnu.org.}{A top-down
  approach to ic design} (2014).
\newline\urlprefix\url{http://www.gnu.org.}

\bibitem{martin2001system}
G.~Martin, H.~Chang, System-on-chip design, in: ASICON 2001. 2001 4th
  International Conference on ASIC Proceedings (Cat. No. 01TH8549), IEEE, 2001,
  pp. 12--17.

\bibitem{kondo2015three}
K.~Kondo, M.~Kada, K.~Takahashi, Three-Dimensional Integration of
  Semiconductors: Processing, Materials, and Applications, Springer, 2015.

\bibitem{sutherland1964sketch}
I.~E. Sutherland, Sketch pad a man-machine graphical communication system, in:
  Proceedings of the SHARE design automation workshop, 1964, pp. 6--329.

\bibitem{vladimirescu1980spice}
A.~Vladimirescu, A.~R. Newton, D.~O. Pederson, SPICE Version 2G. 1 user's
  Guide, University of California Berkeley, Ca, 1980.

\bibitem{shiva1979computer}
S.~G. Shiva, Computer hardware description languages—a tutorial, Proceedings
  of the IEEE 67~(12) (1979) 1605--1615.

\bibitem{kuon2008fpga}
I.~Kuon, R.~Tessier, J.~Rose, et~al., Fpga architecture: Survey and challenges,
  Foundations and Trends{\textregistered} in Electronic Design Automation 2~(2)
  (2008) 135--253.

\bibitem{LEFDEF}
I.~Cadence Design~Systems, Lef/def 5.8 language reference,
  \url{http://coriolis.lip6.fr/doc/lefdef/lefdefref/LEFSyntax.html} (2017).

\bibitem{gajski2012high}
D.~D. Gajski, N.~D. Dutt, A.~C. Wu, S.~Y. Lin, High—Level Synthesis:
  Introduction to Chip and System Design, Springer Science \& Business Media,
  2012.

\bibitem{huang2021machine}
G.~Huang, J.~Hu, Y.~He, J.~Liu, M.~Ma, Z.~Shen, J.~Wu, Y.~Xu, H.~Zhang,
  K.~Zhong, et~al., Machine learning for electronic design automation: A
  survey, ACM Transactions on Design Automation of Electronic Systems (TODAES)
  26~(5) (2021) 1--46.

\bibitem{sivacadence}
C.~Siva, Cadence cloud—the future of electronic design automation.

\bibitem{drechsler2004advanced}
R.~Drechsler, et~al., Advanced formal verification, Vol. 122, Springer, 2004.

\bibitem{sherwani2012algorithms}
N.~A. Sherwani, Algorithms for VLSI physical design automation, Springer
  Science \& Business Media, 2012.

\bibitem{lod_AIA}
A.~I. of~Architects, Aia document e202tm-2008 building information modeling
  protocol exhibit, American Institute of Architects.

\bibitem{Hong_buildsim_challenges}
T.~Hong, J.~Langevin, K.~Sun,
  \href{https://doi.org/10.1007/s12273-018-0444-x}{Building simulation: Ten
  challenges}, Building Simulation 11~(5) (2018) 871--898.
\newblock \href {http://dx.doi.org/10.1007/s12273-018-0444-x}
  {\path{doi:10.1007/s12273-018-0444-x}}.
\newline\urlprefix\url{https://doi.org/10.1007/s12273-018-0444-x}

\bibitem{IFC_iso201816739}
I.~ISO, 16739-1: 2018: Industry foundation classes (ifc) for data sharing in
  the construction and facility management industries—part 1: Data schema,
  International Organisation for Standardisation: Geneva, Switzerland.

\bibitem{gbxml}
gbXML, Green building xml schema, \url{https://www.gbxml.org/} (2000).

\bibitem{hietanen2006ifc}
J.~Hietanen, S.~Final, Ifc model view definition format, International Alliance
  for Interoperability (2006) 1--29.

\bibitem{froese2003future}
T.~Froese, Future directions for ifc-based interoperability, Journal of
  Information Technology in Construction (ITcon) 8~(17) (2003) 231--246.

\bibitem{zhao2019real}
J.~Zhao, O.~Sepp{\"a}nen, A.~Peltokorpi, B.~Badihi, H.~Olivieri, Real-time
  resource tracking for analyzing value-adding time in construction, Automation
  in Construction 104 (2019) 52--65.

\bibitem{holm2021construction}
L.~Holm, J.~E. Schaufelberger, Construction cost estimating, Routledge, 2021.

\bibitem{cavieres2018functional}
A.~Cavieres, A functional modeling framework for interdisciplinary building
  design, Ph.D. thesis, Georgia Institute of Technology (2018).

\end{thebibliography}
 
\end{document}